\documentclass[aps,twocolumn,amsmath,amsfonts,nofootinbib,preprintnumbers,
  superscriptaddress,eqsecnum,secnumarabic]{revtex4-1}
\usepackage[utf8]{inputenc} 
\usepackage{color,graphicx,hyperref,amsmath,amssymb,multirow}
\hypersetup{
   bookmarks=true,         
   unicode=true,          
   pdftoolbar=true,        
   pdfmenubar=true,        
   pdffitwindow=false,     
   pdfstartview={FitH},    
   pdftitle={DFG},    
   pdfauthor={Author},     
   pdfsubject={Subject},   
   pdfcreator={Creator},   
   pdfproducer={Producer}, 
   pdfkeywords={keyword1} {key2} {key3}, 
   pdfnewwindow=true,      
   colorlinks=true,       
   linkcolor=blue,          
   citecolor=magenta,        
   filecolor=magenta,      
   urlcolor=cyan           
}

\newcommand{\be}{\begin{equation}}
\newcommand{\ee}{\end{equation}}
\def\bsp#1\esp{\begin{split}#1\end{split}}
\newcommand{\ie}{\textit{i.e.}}

\newcommand{\amc}{{\sc MadGraph5}\_a{\sc MC@NLO}}
\newcommand{\fr}{{\sc Feyn\-Rules}}
\newcommand{\fa}{{\sc Feyn\-Arts}}
\newcommand{\nloct}{{\sc NloCT}}

\newcommand{\mspin}{{\sc MadSpin}}
\newcommand{\mw}{{\sc MadWidth}}
\newcommand{\ma}{{\sc MadAnalysis~5}}
\newcommand{\py}{{\sc Pythia~8}}
\newcommand{\del}{{\sc Delphes~3}}
\newcommand{\fj}{{\sc FastJet}}

\def\so{O_R}
\def\pso{O_I}

\def\ggo{g_8}
\def\gqL{{y_{8}^L}}
\def\gqR{{y_{8}^R}}

\newcommand{\scr}[1]{\ensuremath{\mathcal{#1}}}
\def\nn{\nonumber}

\begin{document}

\title{Cornering sgluons with four-top-quark events}
\author{Luc Darm\'e}
\email{luc.darme@ncbj.gov.pl}
\affiliation{National Centre for Nuclear Research,
  Ho{\. z}a 69, 00-681 Warsaw, Poland}
\author{Benjamin Fuks}
\email{fuks@lpthe.jussieu.fr}
\affiliation{Laboratoire de Physique Th\'eorique et Hautes Energies (LPTHE), UMR 7589,
Sorbonne Universit\'e et CNRS, 4 place Jussieu, 75252 Paris Cedex 05, France}
\affiliation{Institut Universitaire de France, 103 boulevard Saint-Michel,
  75005 Paris, France}
\author{Mark Goodsell}
\email{goodsell@lpthe.jussieu.fr}
\affiliation{Laboratoire de Physique Th\'eorique et Hautes Energies (LPTHE), UMR 7589,
Sorbonne Universit\'e et CNRS, 4 place Jussieu, 75252 Paris Cedex 05, France}

\date{\today}

\begin{abstract}
The existence of colour-octet scalar states, often dubbed sgluons, is predicted
in many extensions of the Standard Model of particle physics, such as supersymmetric realisations featuring Dirac gauginos.
Such states have a large pair-production rate at hadron colliders and mainly
decay into pairs of jets and top quarks.
Consequently, they
represent a primary target for experimental
searches for new resonances in the multijet and multitop channels at the Large
Hadron Collider. Adopting a phenomenologically-motivated simplified model, we
reinterpret the results of a recent experimental search for the four-top-quark Standard Model signal,
 from which we constrain the sgluon mass to be larger
 than about 1.06~TeV.
We additionally consider how modifications of the existing
four-top-quark studies could enhance our ability to unravel the presence of
scalar octets 
 in data.
\end{abstract}

\maketitle

\section{Introduction}

Since its discovery as the heaviest particle of the Standard Model of particle
physics, the top quark is considered as an attractive probe for new physics, in
particular at the Large Hadron Collider (LHC). In many theories extending the
Standard Model, top-quark production at the LHC is indeed expected to be
enhanced by contributions originating from the decay of new states into one or
more top quarks. Moreover, as the corresponding Standard
Model background is usually well understood, top-quark studies consist of
particularly clean targets for new physics searches, especially when leptonic
top decays are in order.

Whilst most studies have so far solely focused on single-top and
top-quark pair production, the increased integrated luminosity and
centre-of-mass energy of the LHC Run~2 reinforce the potential role of
new physics probes exhibiting a higher top-quark multiplicity. Among all
interesting processes, there has been a growing experimental push in measuring
the four-top production cross section at the LHC. The most recent experimental
measurement undertaken by the CMS collaboration~\cite{Sirunyan:2017roi},
\be
  \sigma_{4t}^{\rm exp}  = 16.9^{+13.8}_{-11.4}~{\rm fb},
\ee
has been found to agree with the Standard Model
expectation~\cite{Frederix:2017wme},
\be
  \sigma_{4t}^{\rm SM} = 11.97^{+2.15}_{-2.51}~{\rm fb} \ ,
\ee
where the latter results include next-to-leading order (NLO) corrections both in
QCD and in the electroweak theory, and where the error only accounts for scale
uncertainties. The theoretical and (still statistically-limited) experimental
precision available today nevertheless open bright prospects for advancing our
knowledge on new physics from four-top probes, as the room for a significant
beyond the Standard Model contribution starts to be more and more reduced.

Many new physics theories feature heavy coloured resonances that essentially
decay into a pair of top quarks, so that their pair production could hence yield
an enhancement of the four-top production cross section. Conversely, the current
measurement of the latter and its confrontation to theoretical predictions in
varied new physics setups could lead to constraints on the corresponding models.
For instance, models with composite top quarks have historically offered one of
the theoretical motivations for LHC searches in this channel~\cite{%
Lillie:2007hd,Gerbush:2007fe,Pomarol:2008bh,Kumar:2009vs} and led to early
recasting results~\cite{Zhou:2012dz}.
In this work,
we focus instead on scenarios featuring the presence of light scalar or pseudoscalar
states lying in the adjoint representation of the QCD gauge group. Such states
are commonly dubbed sgluons and for instance arise in non-minimal supersymmetric
models~\cite{Salam:1974xa,Fayet:1974pd,Fayet:1975yi,AlvarezGaume:1996mv,Fox:2002bu,%
Kribs:2007ac,Benakli:2008pg,Benakli:2014cia,Goodsell:2014dia,Benakli:2016ybe,Benakli:2018vqz}, vector-like confining theories~\cite{Kilic:2009mi} and
extra-dimensional frameworks~\cite{Burdman:2006gy}. 
Among those models, the
supersymmetric case is especially interesting as it predicts the existence of
a \emph{complex} colour-octet scalar which is split into \emph{two} non-degenerate real components after supersymmetry-breaking. 
When its couplings preserve CP these become a real scalar and a real pseudoscalar. The pseudoscalar
state is generally expected to be lighter, as unrelated to the heavy gluino field,
contrary to its scalar sibling, and it solely decays into a pair of quarks where
the top-antitop channel dominates. In contrast, the scalar resonance is
generally heavier and decays both into quarks and gluons. Investigating sgluon
production and decay, both in the scalar and pseudoscalar cases, is therefore
crucial for assessing the viability of this class of supersymmetric extensions.
In this context, four-top production can be considered as one of the most
relevant smoking guns as it could receive contributions from the lightest new
physics particles.

Pioneering studies have been dedicated to the phenomenology of both
complex~\cite{Plehn:2008ae,Choi:2008ub,GoncalvesNetto:2012nt} and real~\cite{%
Schumann:2011ji,Calvet:2012rk,Degrande:2014sta,Goodsell:2014dia} sgluons. More recently,
the results of the LHC Run~1 have been confronted to the predictions of a
simplified real sgluon model~\cite{Calvet:2012rk}, using state-of-the-art Monte
Carlo simulations matching fixed-order calculations at the next-to-leading order
in QCD with parton showers~\cite{Degrande:2014sta} and relying on several LHC
four-top studies~\cite{Beck:2015cga}. The advantage of such an approach is that
bounds can easily be reinterpreted in different, possibly ultraviolet-complete,
theoretical frameworks~\cite{Valencia:2016npc,Benakli:2016ybe,%
BuarqueFranzosi:2017qlm,Hayreter:2017wra}. On the other hand, early four-top LHC
Run~2 results from
the ATLAS collaboration~\cite{Aaboud:2017vwy} have also been reinterpreted, this
time to constrain a pseudoscalar sgluon model~\cite{Kotlarski:2016zhv}.

Driven by the recent experimental progress on the path to a potential four-top
signal observation~\cite{Sirunyan:2017roi}, we investigate in this letter how
the analysis of 35.9~fb$^{-1}$ of LHC proton-proton collisions at a
centre-of-mass energy of 13~TeV could constrain light pseudoscalar sgluons, such
as those predicted in supersymmetric models featuring Dirac gauginos. To this
aim, we have implemented the above-mentioned CMS four-top study in the \ma\
framework~\cite{Conte:2012fm,Conte:2014zja} and made it publicly available
through the \ma\ Public Analysis Database~\cite{Dumont:2014tja} and
{\sc InSpire}~\cite{1672876}. While this offers a possibility to test any given
new physics model against the results of this particular search, we focus on the
pseudoscalar sgluon case and use state-of-the-art Monte Carlo
simulations~\cite{Degrande:2014sta} to
constrain the associated four-top signal. We moreover present some of the most
interesting features that could allow for the distinction of Standard Model and
sgluon-induced four-top production.

The rest of this letter is organised as follows. We briefly review the
simplified sgluon model that we adopt in Sec.~\ref{sec:model} and connect it to
one of its possible ultraviolet origins. We next detail in Sec.~\ref{sec:recast}
our reimplementation of the considered four-top CMS analysis and reinterpret its
results in the context of our sgluon simplified model. Sec.~\ref{sec:kinematics}
is then dedicated to the presentation of various handles on sgluon-induced
four-top production that could be enhance the LHC sensitivity to these states in
the future. We finally summarise our findings in Sec.~\ref{sec:summary}.

\section{Simplified model of sgluons}
\label{sec:model}

Our phenomenological analysis relies on a simplified sgluon model in which the
Standard Model is supplemented by a scalar colour-octet field $O$ of mass $m_O$,
singlet under the electroweak gauge group and traditionally dubbed sgluon.
After electroweak symmetry breaking, 
the Lagrangian describing its dynamics is given by~\cite{Calvet:2012rk}
\begin{align}
\scr{L} = &
       \frac12 D_\mu O^a D^\mu O^a \!-\! \frac12 m_O^2 O^a O^a \nn\\
 &    \!+\! \ggo d_{abc} O^a G_{\mu \nu}^b G^{\mu \nu c}\! + \tilde{g}_8 d_{abc} O^a G_{\mu \nu}^b \tilde{G}^{\mu \nu c} \nn\\
& \!+\! \Big\{ \bar{q} \Big[ {\bf \gqL} P_L + {\bf \gqR} P_R \Big] O^a T^a q +
      {\rm  h.c.} \Big\} \ ,  \label{eq:Lag_eff} 
\end{align}
where fundamental colour and flavour indices are understood for clarity, $T^a$
and $d_{abc}$ respectively stand for the fundamental representation matrices and
symmetric structure constants of $SU(3)$, $P_{L,R}$ are the usual chirality
projectors, and $G_{\mu \nu}^a$ ($\tilde{G}_{\mu \nu}^{a}$) is the gluon field strength (dual field strength). This Lagrangian includes standard gauge-invariant kinetic and mass
terms for a real field lying in the adjoint representation of the QCD gauge
group, as well as the effective interactions of a single sgluon with
the Standard Model quarks and gluons. The respective interaction strengths of
the latter are embedded within the ${\bf y_8^{L,R}}$ matrix parameters in
generation space and the (dimensionful) $g_8$ parameter. 

Since the octets are real fields, the couplings must satisfy ${\bf y^L_8} =
{\bf y^R_8}^\dag$. Furthermore, we assume in the following that $CP$ is
conserved in the new physics sector, so that we can distinguish a scalar from a
pseudoscalar sgluon.
For a
pseudoscalar sgluon $g_8$ vanishes and the ${\bf y_8}$
matrices are pure imaginary; whereas in the scalar case, the fermion couplings are real and $\tilde{g}_8$ vanishes. 

As a precise example stemming from a top-down perspective, one can link our
simplified model of Eq.~\eqref{eq:Lag_eff} to a class of supersymmetric
realisations featuring Dirac gauginos. The spectrum of such models typically
contains a complex scalar octet field $\mathbf{O}$ whose scalar ($\so$) and
pseudoscalar ($\pso$) component obtain different masses after supersymmetry
breaking. Introducing the supersymmetry-breaking scalar-octet mass $m_O$
and bilinear coupling $B_O$, as well as the Dirac gluino mass $m_D$, the
mass terms of the $\so$ and $\pso$ components read~\cite{Goodsell:2014dia}
\be\bsp
 \hspace*{-.2cm}{\cal L}_O = &\
    -\! m_O^2 |\mathbf{O}|^2
    \!-\! \frac12 ( B_O \mathbf{O}^2 \!+\! {\rm h.c.})
    \!-\! (m_D \mathbf{O} \!+\! {\rm h.c.})^2\\
  \underset{CP}{=}&\ -\! \frac12 ( m_O^2 \!+\! B_O \!+\! 4 m_D) \so^2
      -\! \frac12 ( m_O^2 \!-\! B_O) \pso^2 \ .
\esp\ee
As the pseudoscalar mass is not related to the gluino mass, 
it can easily lie in the sub-TeV
mass range while respecting the severe gluino LHC
bounds~\cite{Benakli:2016ybe}; moreover, that it should be lighter than the stops and other SUSY particles is a \emph{prediction} of the ``Goldstone gaugino'' scenario \cite{Alves:2015kia,Alves:2015bba}. Finally, in Dirac gaugino models
\begin{equation}
\tilde{g}_8 = 0
\end{equation}
to one loop order \cite{Goodsell:2014dia}, while the  ${\bf y_8^{L,R}}$ couplings are generated at one loop and are proportional to the masses of the respective quarks.
This last property is generic and stems from the fact that the
quark-quark-sgluon interactions of Eq.~\eqref{eq:Lag_eff} are not
$SU(2)$-invariant and that any new flavour-violating effects cannot be
significantly larger than in the Standard Model.
Realistic sgluon benchmark configurations
could hence feature, motivated by this top-down approach, a relatively heavy
scalar sgluon coupling both to gluons and quarks and a relatively light
pseudoscalar sgluon coupling almost entirely to top quarks with a small coupling to bottom quarks.

\begin{figure}
 \begin{center}
  \includegraphics[width=0.25\textwidth]{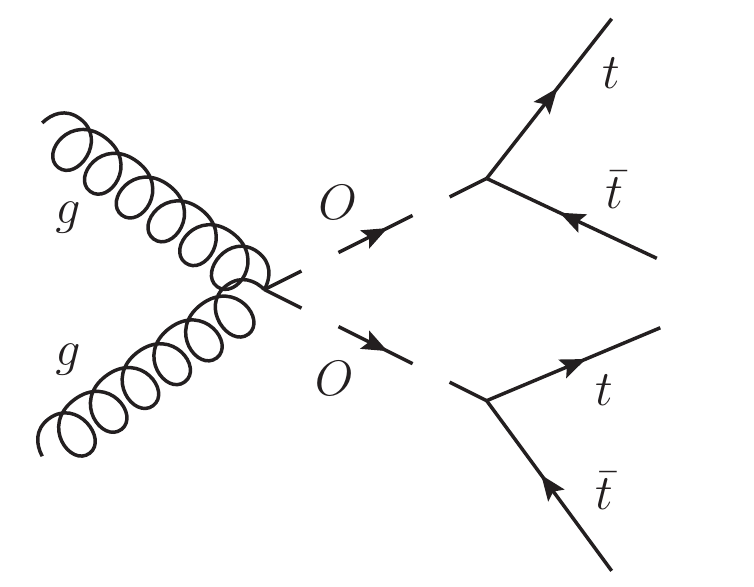}
 \end{center}
 \vspace*{-.6cm}
 \caption{Representative Feynman diagram illustrating sgluon pair production and
   decay into a four-top system.}
 \label{fig:4tprod}
\end{figure}

{\setlength{\tabcolsep}{0.4em}
 \begin{table*}
  \centering\footnotesize
  \begin{tabular}{|c|p{4cm}|p{4cm}|p{3cm}|p{3cm}|}
   \hline \hline
   \multicolumn{5}{|l|}{\textbf{Object reconstruction}} \\
   \hline
   & Electrons & Muons & Jets & $b$-tagged jets \\
   $p_T$ (GeV) & $>20$ &  $>20$ &  $>40$ &  $>25$ \\
   $\eta$ & $<2.5$ &  $<2.4$ &  $<2.4$ &  $<2.4$ \\
   \hline
   \multicolumn{5}{|l|}{\textbf{Isolation}: All jets used for imposing lepton
    isolation are discarded~\cite{Khachatryan:2016kod}.} \\
   \hline \hline
   \multicolumn{5}{|l|}{\textbf{Baseline selection}} \\
   \hline
   Jets & \multicolumn{4}{|c|}{$H_T > 300$ GeV,  $p_T^{\text{miss}} > 50$ GeV,
     at least two jets and two $b$-tagged jets.} \\
   Leptons & \multicolumn{4}{|c|}{Pair of same-sign isolated leptons, with the
     leading one satisfying $p_T > 25$ GeV.} \\
   Vetoes & \multicolumn{4}{|p{14cm}|}{Third loosely-isolated electron (muon)
     with $p_T > 5~(7)$ GeV forming an opposite-sign same-flavour lepton pair
     with an invariant mass $m_{\text{OS}} < 12$ GeV or $m_{\text{OS}} \in
    [76,106]$ GeV.} \\
   \hline
  \end{tabular}
  \caption{Summary of the object reconstruction and baseline selection procedure
    of the CMS four-top analysis of Ref.~\cite{Sirunyan:2017roi}.}
  \label{Tab:CMScuts}
 \end{table*}
}

A generic prediction of Dirac gaugino models is then that the pseudoscalar sgluons should decay almost exclusively to tops. If pseudoscalar sgluons are indeed light, they are expected to be copiously produced at hadronic colliders, the four-top signal arising then from 
their decay into a top-antitop pair with an almost 100\% branching ratio (see
Fig.~\ref{fig:4tprod}). In this work, we compare predictions for this
sgluon-induced four-top signal with the recent measurement achieved by the CMS
collaboration in the multileptonic channel~\cite{Sirunyan:2017roi}, and 
suggest that a more dedicated search strategy is necessary to potentially improve the limits in
the upcoming years. To this aim, we have implemented a common
scalar and pseudoscalar sgluon simplified model in \fr~\cite{Alloul:2013bka},
which we jointly use with \nloct~\cite{Degrande:2014vpa} and
\fa~\cite{Hahn:2000kx} to generate a UFO module~\cite{Degrande:2011ua} allowing
for NLO calculations in QCD within the \amc\ framework~\cite{Alwall:2014hca}.

The simulation of the sgluon signal is achieved by generating hard scattering
events where NLO matrix elements in QCD are convoluted with the NLO set of
NNPDF3.0 parton densities~\cite{Ball:2014uwa}. After including sgluon decays
into a top-antitop system as performed by \mspin~\cite{Artoisenet:2012st} and
\mw~\cite{Alwall:2014bza}, the fixed-order results are matched
with parton showers, the latter being described by \py~\cite{Sjostrand:2014zea}
that also takes care of the simulation of the hadronisation effects. We finally
model the response of the CMS detector with \del~\cite{deFavereau:2013fsa}, that
internally relies on \fj~\cite{Cacciari:2011ma} for object reconstruction, and
we mimic the CMS four-top selection strategy by reimplementing the analysis of
Ref.~\cite{Sirunyan:2017roi} in the \ma~\cite{Conte:2012fm,Conte:2014zja,%
Dumont:2014tja} framework.

\section{Pseudoscalar sgluon bounds from four-top production}
\label{sec:recast}
In the scenarios under consideration, pseudoscalar sgluons almost exclusively couple to top quarks, so that all
existing sgluon bounds are automatically evaded. The latter are indeed derived
from resonance searches in top-antitop~\cite{Aaboud:2018mjh,%
Sirunyan:2017uhk} or dijet~\cite{Aaboud:2017yvp,CMS:2017mog,CMS:2017xrr} final
states, from the shape of the $t\bar t$ differential cross section~\cite{%
Aaboud:2017fha,Aaboud:2018eqg,Khachatryan:2016mnb,Sirunyan:2017mzl} or from
dijet pair production~\cite{Aaboud:2017nmi,CMS:2016pkl}, which all require a
non-vanishing sgluon coupling either to light quarks or to gluons or to both.
The only potential constraints hence arise from searches for new physics in the
four-top final state. For this reason, we have implemented the most recent
Standard Model CMS four-top analysis~\cite{Sirunyan:2017roi} in the \ma\
framework, and we have used our reimplementation to revisit the CMS results in
the context of our pseudoscalar sgluon model. Our {\sc C++} code is additionally
publicly available from {\sc InSpire}~\cite{1672876} through its digital object
identifier (DOI) \href{http://doi.org/10.7484/INSPIREHEP.DATA.BBBC.6732}%
{10.7484/INSPIREHEP.DATA.BBBC.6732}.

The main object reconstruction features and baseline selection criteria of the
considered search are summarised in Table~\ref{Tab:CMScuts}. Jets are
reconstructed by means of the anti-$k_T$ algorithm~\cite{Cacciari:2008gp} with a
radius parameter $R=0.4$, and only those jets with a transverse momentum $p_T$
and pseudorapidity $\eta$ satisfying the criteria indicated in the top panel of
the table are retained, a slighter selection being imposed on $b$-tagged jets.
Furthermore, electron and muon candidates are required to be central and rather
hard (see again the top panel of the table), and the various reconstructed
objects are imposed to be isolated as described in Ref.~\cite{%
Khachatryan:2016kod}. The baseline selection first includes constraints on the
hadronic activity $H_T$, defined as the scalar sum of the $p_T$ of all
reconstructed jets, and on the missing transverse energy $p_T^{\rm miss}$. It
next requests the presence of at least two jets, two $b$-tagged jets and one
pair of same-sign leptons. Moreover, events exhibiting a third lepton forming an
opposite-sign same-flavour pair compatible with a low-mass hadronic
resonance or a $Z$-boson are vetoed (see the lower panel of the table).

\begin{figure*}
	\centering
	\includegraphics[width=0.45\textwidth]{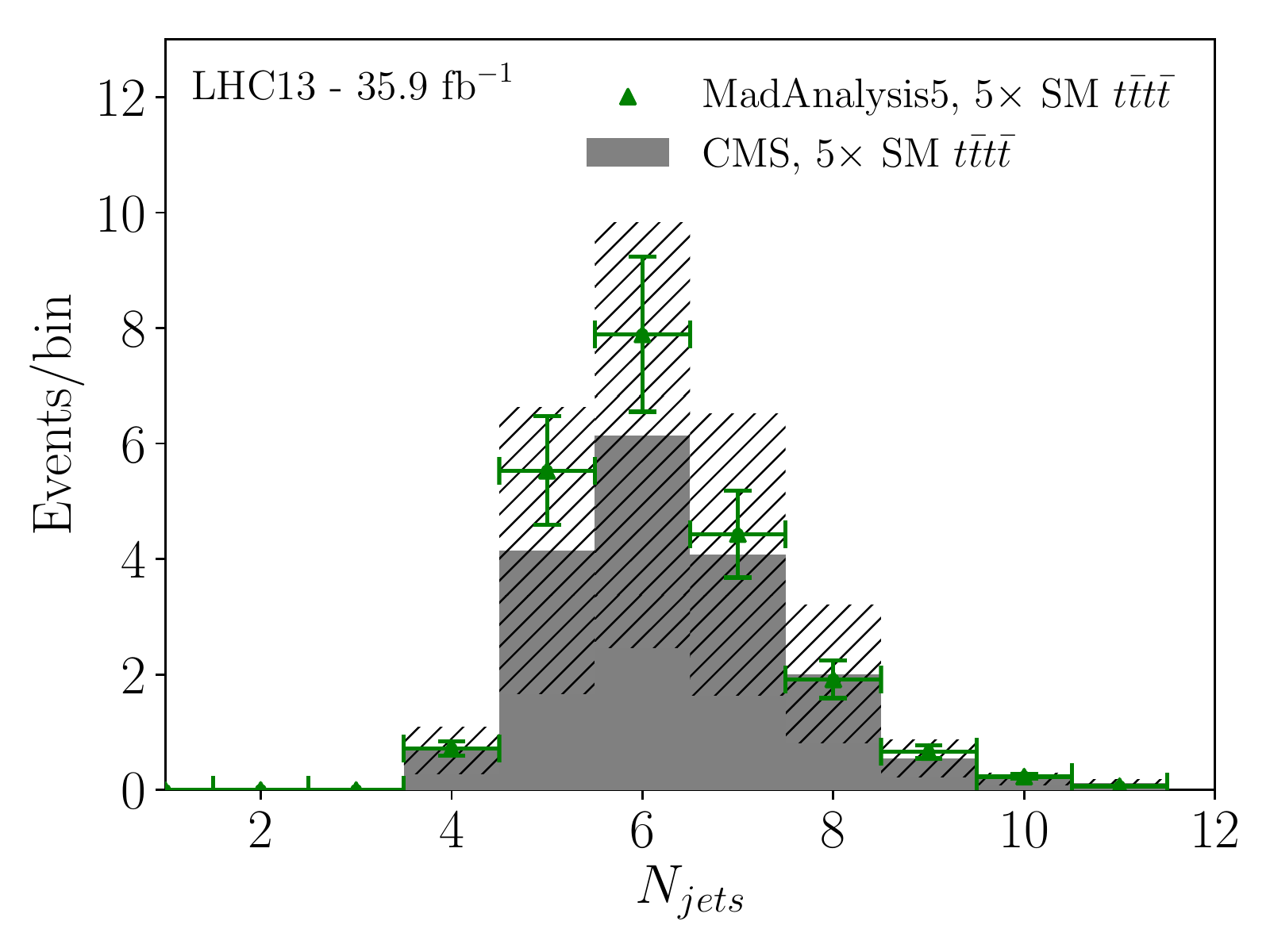}
	\includegraphics[width=0.45\textwidth]{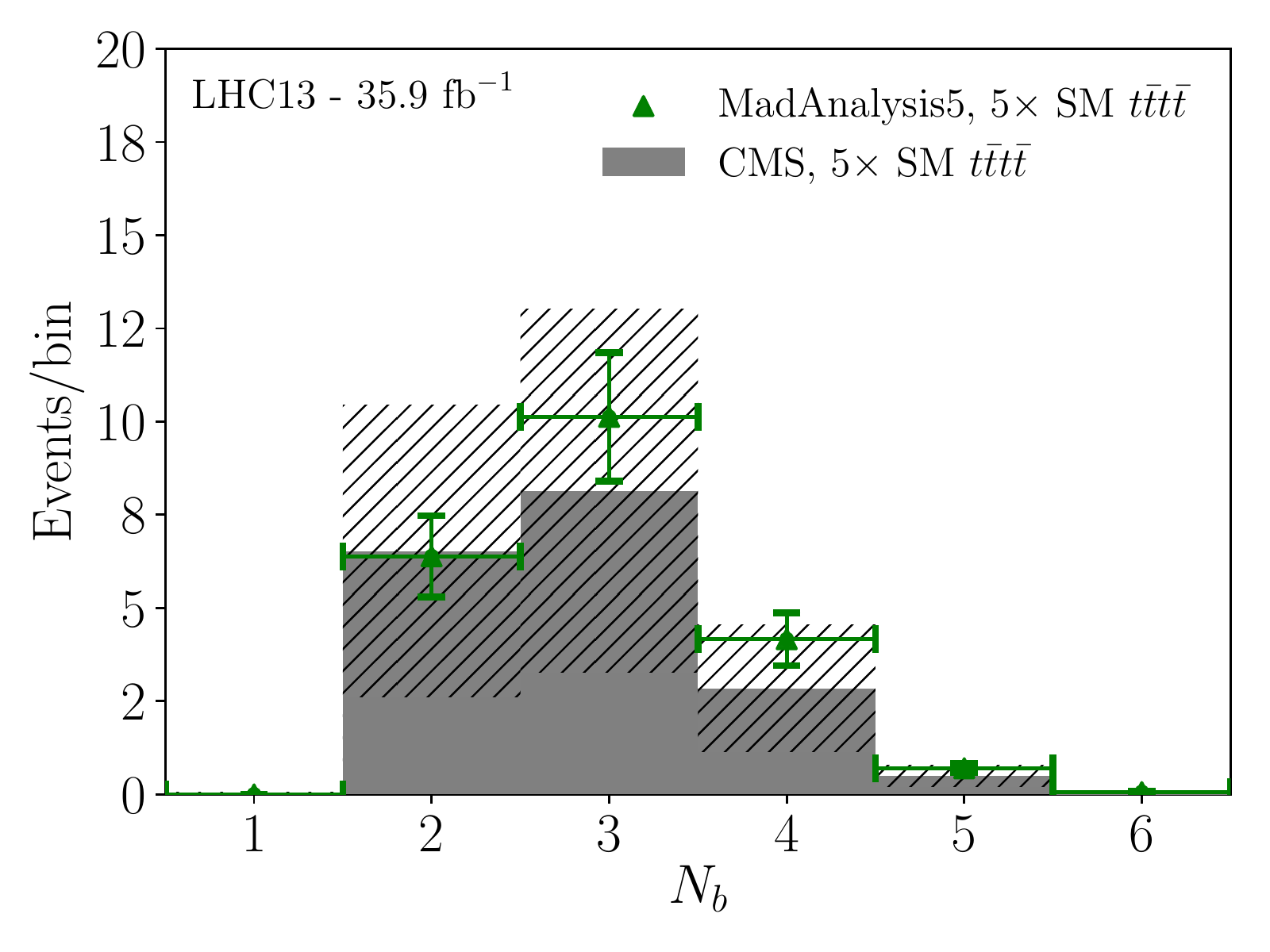}\\
	\includegraphics[width=0.45\textwidth]{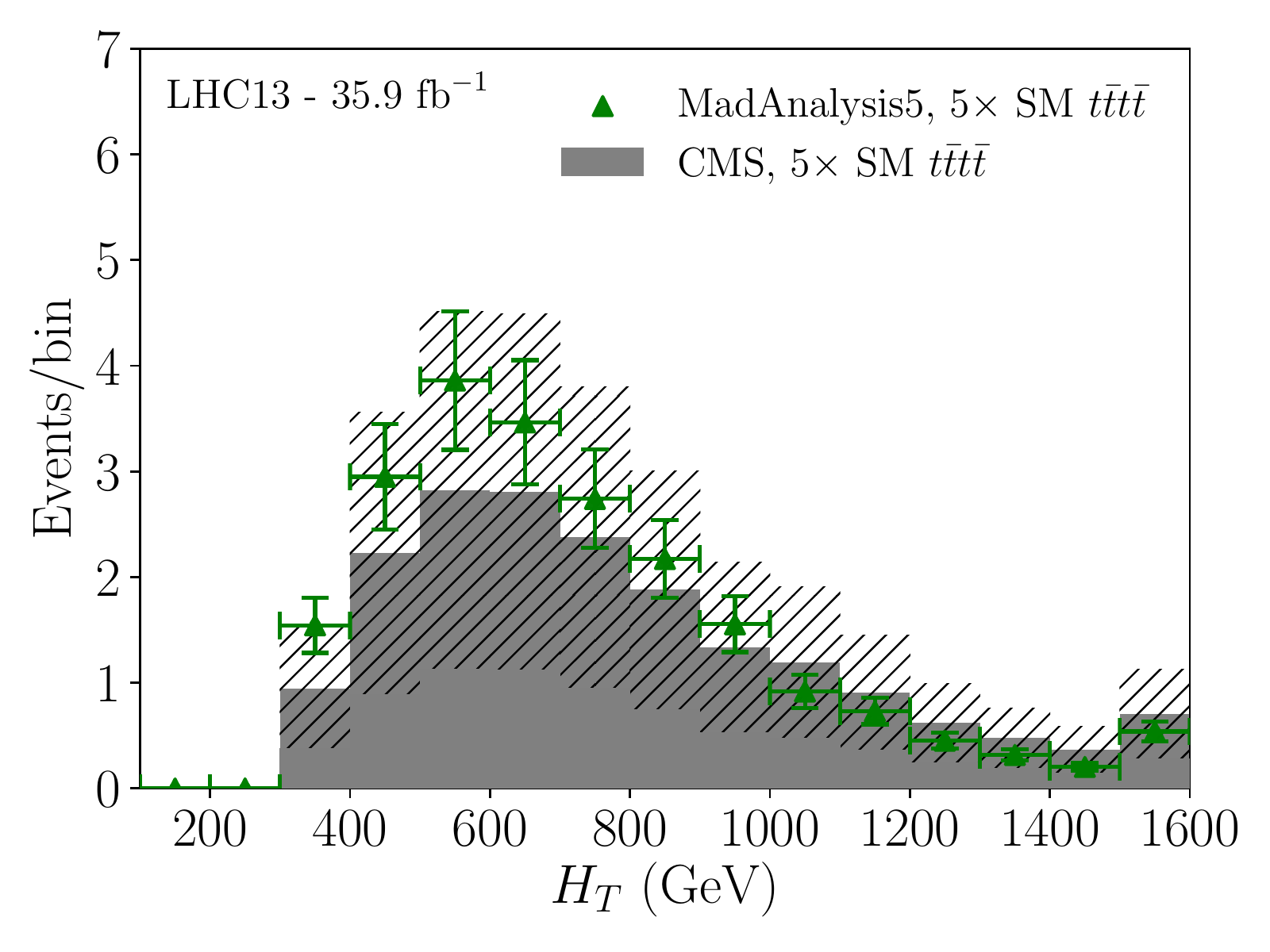}
	\includegraphics[width=0.45\textwidth]{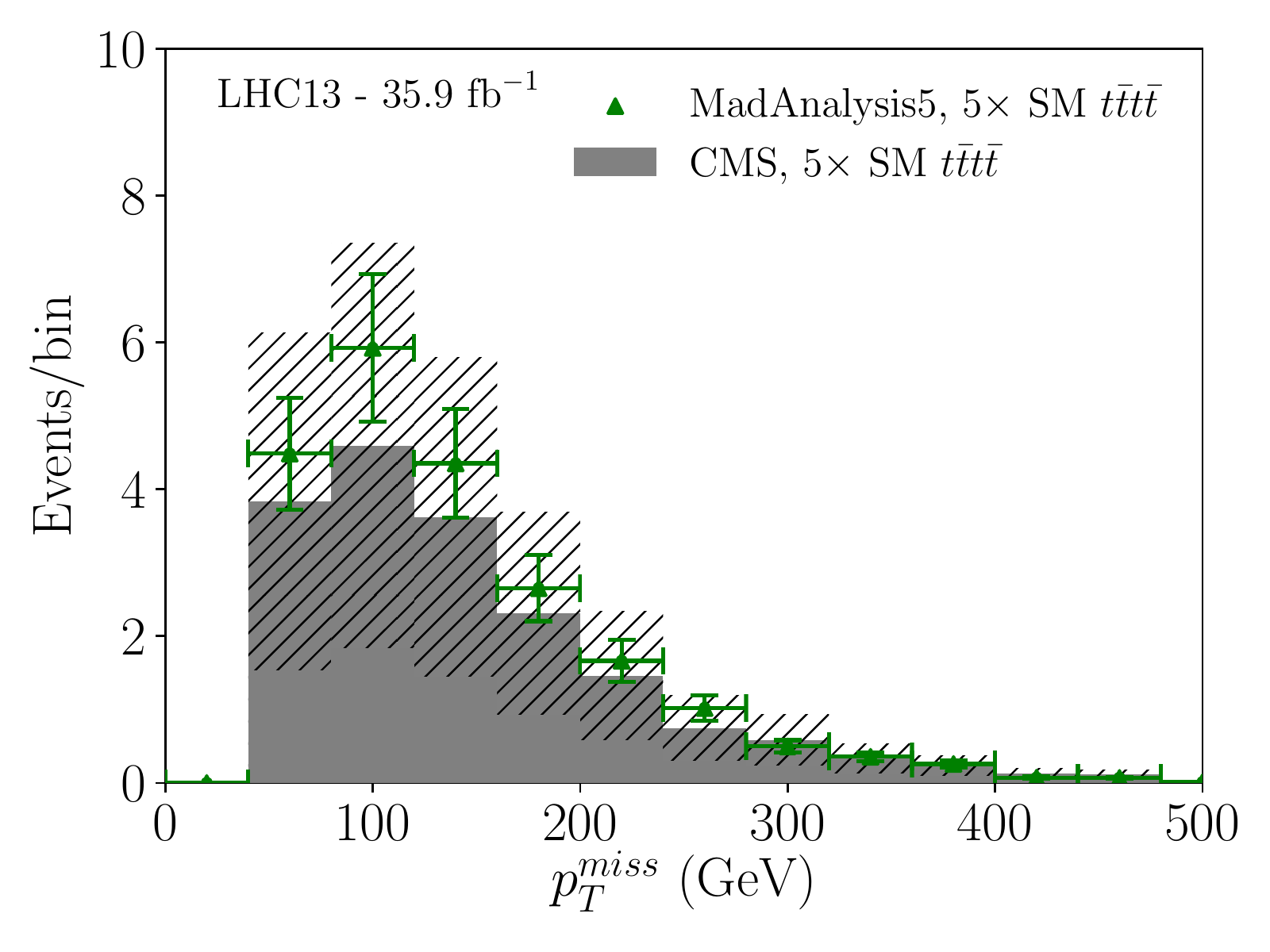}\\
	\includegraphics[width=0.65\textwidth]{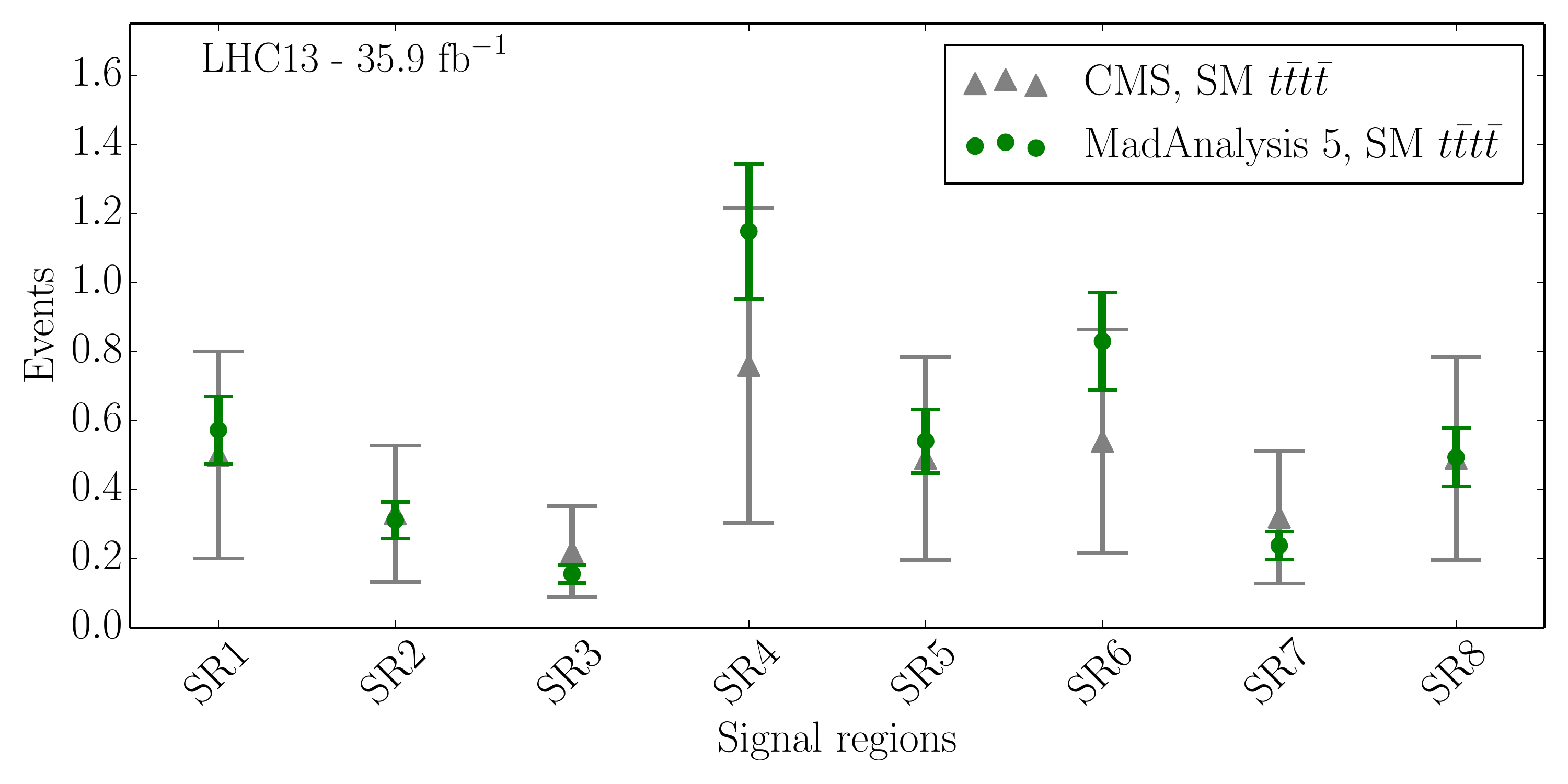}
	\caption{Validation figures of our reimplementation, in the \ma\ framework, of
		the CMS four-top analysis of Ref.~\cite{Sirunyan:2017roi}. We compare \ma\
		predictions (green) with the CMS official results (dark grey) for the jet
		multiplicity (upper-left panel), $b$-jet multiplicity (upper-right panel),
		$H_T$ (central-left panel) and $p_T^{\rm miss}$ (central-right panel)
		spectra, as well as for the event counts populating each signal region (lower
		panel). The \ma\ predictions include statistical uncertainties (green error
		bars) whilst the CMS numbers include both systematical and statistical errors
		(black dashed bands and light grey error bars in the lower panel).}
	\label{fig:validation}
\end{figure*}

As is customary in this type of analysis, the selection strategy then relies on
eight non-overlapping signal regions whose definition asks for different
requirements on the lepton and jet multiplicities. A first set of three signal
regions SR1, SR2 and SR3 focuses on events featuring a same-sign dilepton, 2
$b$-tagged jets and respectively 6, 7 and at least 8 jets. The SR4 and SR5
regions are dedicated to events with a same-sign dilepton, 3 $b$-jets and
5--6 or at least 7 jets respectively, whilst the SR6 region allows instead for
at least 4 $b$-jets and at least 5 jets. Finally, two extra regions
concern events with at least 3 leptons, and either 2 $b$-jets and at
least 5 jets (SR7) or at least 3 $b$-jets and at least 4 jets (SR8).

In order to validate our reimplementation, we consider Standard
Model four-top production. We compare \ma\ predictions for various
differential distributions and for the number of events populating each of the
eight signal regions of the considered analysis with official CMS simulated
results. Although a deeper
comparison would have been desirable, for instance by analysing
cutflows for each signal region on a cut-by-cut basis, the necessary validation
material has not been made publicly available by the CMS collaboration. We thus
generate a Standard Model four-top signal using our simulation chain and show,
after imposing the baseline selection, the jet
multiplicity (upper-left panel), $b$-jet multiplicity (upper-right panel), $H_T$
(left central panel) and $p_T^{\rm miss}$ (right central panel) spectra in
Fig.~\ref{fig:validation}. On each
subfigure, we compare our predictions (green) with the CMS official simulation
results (dark grey), including statistical uncertainties for what concerns our
predictions (green error bars) and both the statistical and systematical
uncertainties for the CMS results (light grey band).
In the lower panel of the figure, we present the number of events expected to
populate each of the eight signal regions after imposing the entire selection,
again comparing the results returned
by our simulation chain (green) with the official CMS expectation (grey). A very
good agreement can be observed, so that we consider our reimplementation as
validated.

\begin{figure}
 \begin{center}
  \includegraphics[width=0.45\textwidth]{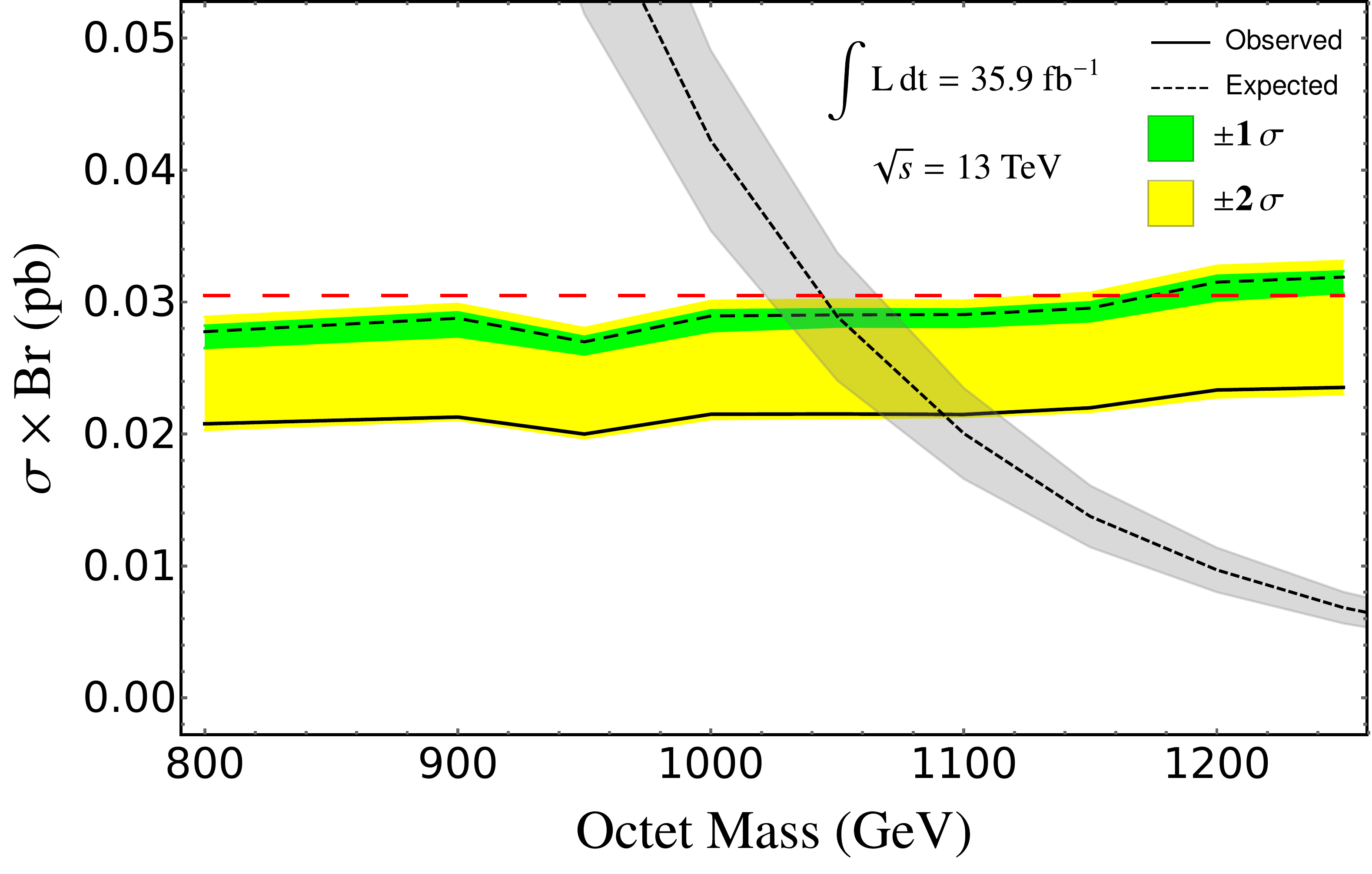}
 \end{center}
 \caption{Expected (dashed) and observed (solid) pseudoscalar sgluon
   pair-production cross section excluded at the 95\%
   confidence level when making use of the results associated with the SR6
   region of the four-top CMS analysis of Ref.~\cite{Sirunyan:2017roi}.
   Theoretical predictions for the signal rate are indicated by the grey band.}
 \label{fig:CS}
\end{figure}

Bounds on our pseudoscalar sgluon model are extracted from the generation of the
corresponding signal for different choices of the sgluon mass $m_O$. For each
signal region of the considered CMS four-top analysis, we evaluate by
means of our simulation chain the number of signal events $n_s$ surviving the
selection, and then confront it to the observed number of events $n_{\rm data}$
after accounting for the Standard Model expectation
$\hat n_b\pm \Delta \hat n_b$. In practice, we generate $10^5$ Monte Carlo toy
experiments in which we take the number of background events $n_b$ from a Gaussian distribution with mean $\hat n_b$
and a width $\Delta \hat n_b$. The $p$-values associated with the
background-only ($p_b$) and signal-plus-background ($p_{s+b}$) hypotheses are
extracted from the Poisson distributions of parameters $n_b$ and $n_b+n_s$
knowing that $n_{\rm data}$ events have been observed. By keeping the signal
total production cross section free, we derive the value for which the new
physics signal is excluded at the 95\% confidence level, \ie\ the smallest cross
section for which
\be
  1 - \frac{p_{s+b}}{p_b} > 0.95 \ .
\ee

The SR6 region is typically the one most populated by the signal, its
selection focusing on one pair of same-sign leptons, at least 4 $b$-jets and at
least 5 hard jets. In Fig.~\ref{fig:CS}, we present the dependence of the cross
section excluded at the 95\% level on the sgluon mass, using the results from
this region only for which
\be
  \hat n_b \pm \Delta \hat n_b = 1.2 \pm 0.4 \quad\text{and}\quad
  n_{\rm data} = 0 \ ,
\label{eq:SR6}\ee
assuming an integrated luminosity of 35.9~fb$^{-1}$. The expected results (when
one considers $n_{\rm data} = n_b$) are shown by
a dashed line, and we include the cross section values spanned by $1\sigma$
(green) and $2\sigma$ (yellow) variations found by repeating the process $10^4$  times, now taking the ``data'' events from a gaussian distribution of mean $\hat{n}_b$ and width $\Delta \hat{n}_b$. The solid line corresponds to the limit from the observed results, which exclude the signal most severely due to the downward fluctuation exhibited in Eq.~\eqref{eq:SR6} (as no event was
observed instead of an expectation of $1.2\pm 0.4$). Conclusive statements are
achieved by superimposing those results to
theoretical predictions for the signal rate extracted from our model. NLO
predictions and the associated theoretical uncertainties are shown by a grey
band, which indicates that pseudoscalar sgluons of mass equal to 1.06~TeV are
conservatively excluded at the 95\% confidence level. We have hence found a
slight gain in exclusion after comparing our results with what could be expected
from the new physics contributions allowed by the CMS four-top total cross
section measurement, the corresponding upper limit being represented by the
red dashed line on the figure.

\section{Dedicated Search Strategy}
\label{sec:kinematics}

\begin{figure*}[!ht]
 \centering
   \includegraphics[width=0.45\textwidth]{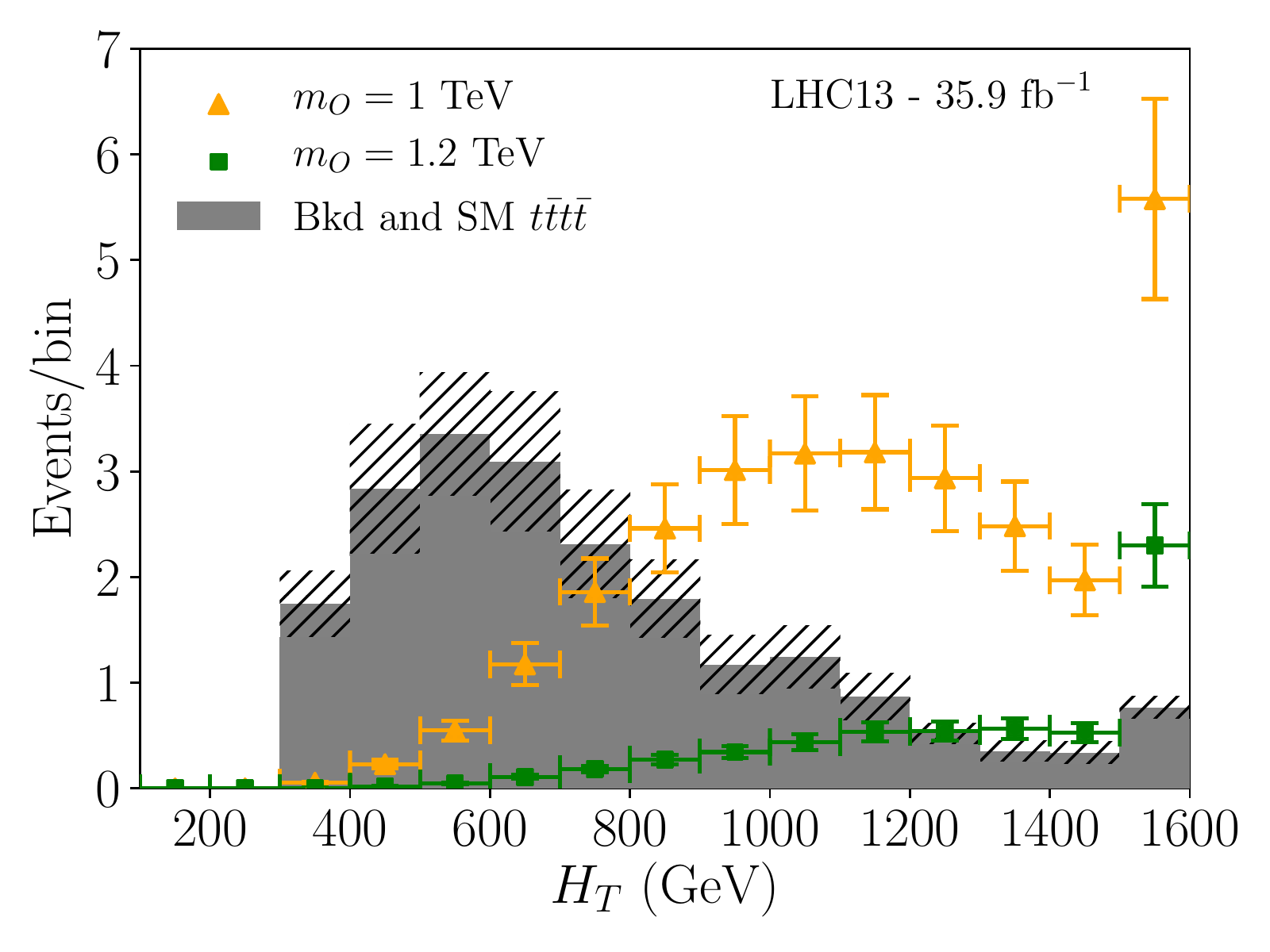}
   \includegraphics[width=0.45\textwidth]{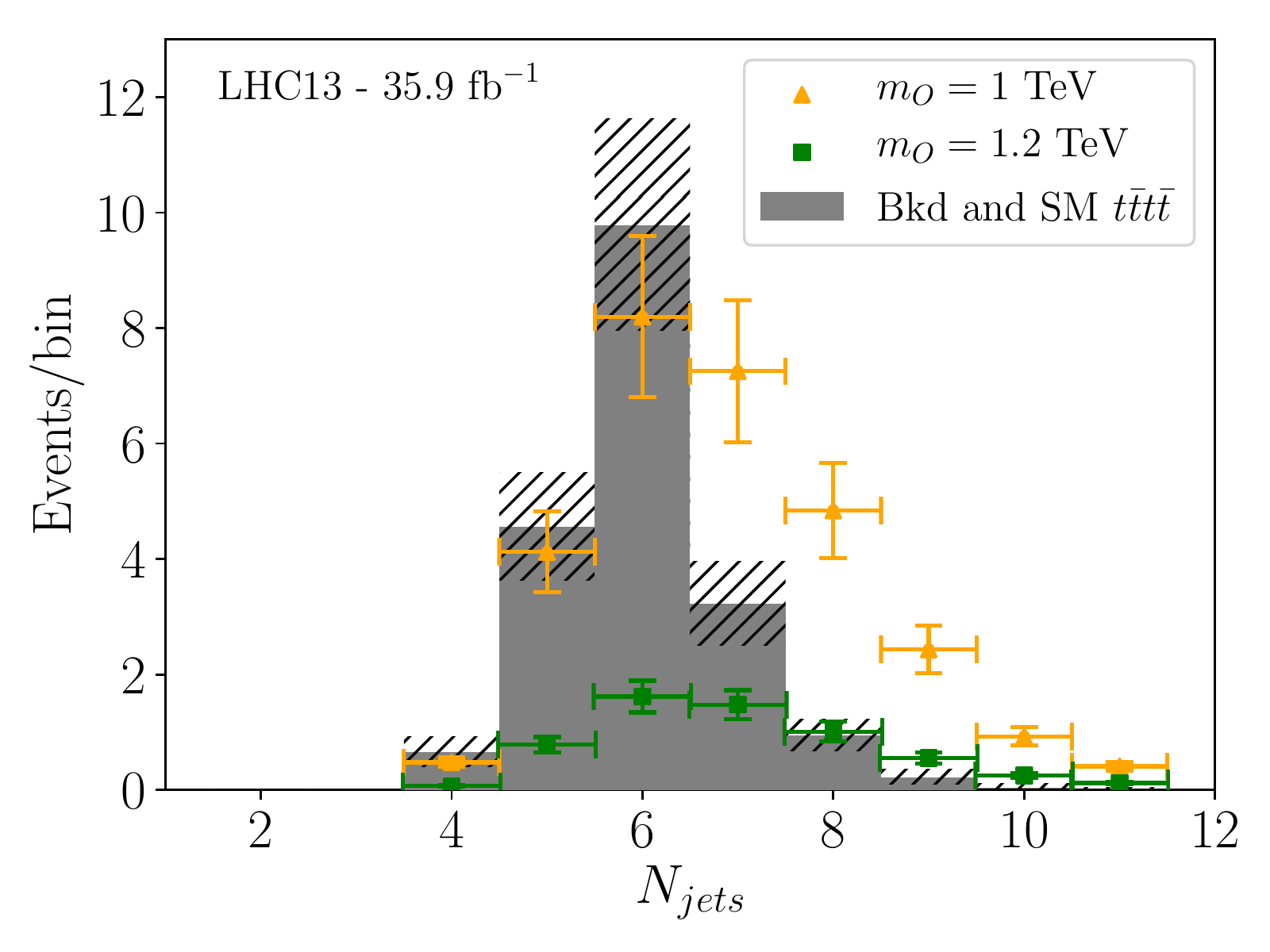}
 \caption{Distributions in the $H_T$ variable (left panel) and in the jet
   multiplicity (right panel) after applying the baseline selection of the CMS
   four-top analysis of Ref.~\cite{Sirunyan:2017roi}. We present the Standard
   Model distributions as provided by the CMS collaboration (dark grey with
   black dashed error bands) and our predictions for the signal in the case of a
   sgluon mass of 1~TeV (orange) and 1.2~TeV (green).}
 \label{fig:KinSR}
\end{figure*}

The CMS analysis under consideration targets the observation of a Standard Model
four-top signal~\cite{Sirunyan:2017roi}. However, the signal selection strategy
is not adapted to a new-physics-induced four-top signal, as the final-state
kinematics could be largely different. For example, we find that the octets are produced almost at rest in the centre-of-mass frame, and the angular distribution of the top decays is therefore flat, in contrast to the SM four-top processes. Consequently, better bounds could be in principle derived from data. In Fig.~\ref{fig:KinSR}, we impose the baseline event selection and present the $H_T$ (upper panel) and jet multiplicity (lower panel) distributions for the background (dark grey with the statistical and systematical uncertainties being encompassed in the light grey band) and two
representative signal scenarios that differ by the value of the sgluon mass
being fixed to $m_O = 1$~TeV (orange) and 1.2~TeV (green). In both cases, we
observe a more important hadronic activity attached to the signal, which stems
from the pair-production and decay of a colour-octet state lying in the TeV mass
range. The signal $H_T$ distribution indeed presents a peak close to the sgluon
mass and it tends to feature a larger jet multiplicity. The $H_T$ variable could
hence provide an excellent discriminant between Standard Model and
sgluon-induced four-top production after imposing a selection cut like
$H_T \gtrsim 800$ GeV. Similarly, relying on probes targeting the tail of the
jet multiplicity spectrum could offer extra handles on the signal, provided that
a sufficient integrated luminosity is available as the statistics steeply falls
with both the sgluon mass and the number of jets. 

\section{Summary}
\label{sec:summary}
In this work, we have investigated how current LHC new physics results constrain
the existence of pseudoscalar fields lying in the adjoint representation of the
QCD gauge group and almost exclusively coupling to top quarks. These fields
naturally arise in many extensions of the Standard Model, and in particular in
supersymmetric realisations featuring Dirac gauginos. By virtue of their
vanishing coupling to light quarks and gluons in these scenarios, pseudoscalar sgluons cannot be probed via standard resonance
searches in the dijet, top-antitop or dijet pair modes, and one must rely
instead on four-top production. Recently, the CMS collaboration has performed
the first measurement of the Standard Model four-top production cross section.
Whilst the error bars are still large and the path to a $5\sigma$ observation is
still long, such a result can already be used to constrain new physics in
general and pseudoscalar sgluons in particular.

We have implemented this CMS search in the \ma\ framework and made it publicly
available for LHC result reinterpretation studies. We have then recast these
results in the context of a pseudoscalar sgluon simplified model. We have shown
that LHC Run~2 results already constrain these sgluons to lie in the TeV mass
regime, the corresponding bound on the sgluon pair-production cross section
being slightly stronger than what could have obtained by using the naive cross-section limit from the four-top production rate measurement. We have
moreover shown that we could benefit from the design of a search dedicated to
a sgluon-induced four-top signal. Specific features in the
hadronic activity associated with a sgluon signal indeed offer a strong
potential in terms of discovery prospects.\\[.5cm]

\acknowledgements

We are grateful to J.~Andrea, F.~Blekman and D.~Lontkovskyi for their help in
understanding some details of the CMS-TOP-17-009 analysis. This work has been
supported in part by French state funds managed by the Agence Nationale de la
Recherche (ANR) in the context of the LABEX ILP (ANR-11-IDEX-0004-02,
ANR-10-LABX-63). MDG acknowledges support from the Agence Nationale de Recherche grant ANR-15-CE31-0002 ``HiggsAutomator.'' LD is supported in part by the National Science Centre (NCN) research grant No.~2015-18-A-ST2-00748. The use of the CIS computer cluster at the National Centre for Nuclear Research in Warsaw is gratefully acknowledged.

\bibliographystyle{utphys}
\bibliography{biblio.bib}
\end{document}